\documentclass[12pt,a4paper]{article}

\usepackage[utf8]{inputenc}
\usepackage[T1]{fontenc}
\usepackage{lmodern}
\usepackage[margin=2.5cm]{geometry}
\usepackage{amsmath,amssymb}
\usepackage{graphicx}
\usepackage{float}
\usepackage{needspace}
\usepackage{booktabs}
\usepackage{array}
\usepackage{longtable}
\usepackage{hyperref}
\usepackage{xcolor}
\usepackage{microtype}
\usepackage{parskip}
\usepackage[numbers,sort&compress]{natbib}
\usepackage{tcolorbox}
\usepackage{enumitem}

\hypersetup{
    colorlinks=true,
    linkcolor=blue!70!black,
    citecolor=blue!70!black,
    urlcolor=blue!70!black,
    pdftitle={What Molecular Structure Cannot Tell Us},
    pdfauthor={Juergen Dietrich},
}

\tcbuselibrary{skins}
\newtcolorbox{infobox}[1]{
    title=#1,
    colback=gray!10,
    colframe=gray!60,
    fonttitle=\bfseries,
    boxrule=0.5pt,
    left=6pt, right=6pt, top=4pt, bottom=4pt
}

\title{\textbf{What Molecular Structure Cannot Tell Us:}\\
A Taxonomy of Explainability Gaps in\\
GNN-Based Drug Toxicity Prediction}

\author{
    Juergen Dietrich\\
    \small ai-solutions-berlin.de\\
    \small \href{mailto:juergen.dietrich@ai-solutions-berlin.de}{juergen.dietrich@ai-solutions-berlin.de}\\
    \small ORCID: \href{https://orcid.org/0000-0002-5494-3499}{0000-0002-5494-3499}
}

\date{May 2026}

\begin{document}
\maketitle
\thispagestyle{empty}

\begin{abstract}
Not all clinically relevant adverse effects are structurally inferable
from molecular graphs --- regardless of model quality or architectural complexity.
This study introduces an operational taxonomy of the structural information limits
that prevent structure-based toxicity prediction, independent of the learning
algorithm employed.
Graph Neural Networks (GNNs) have emerged as a natural approach for molecular
toxicity prediction, operating directly on atomic connectivity without the
information loss inherent to fixed-length fingerprints.
However, the fraction of a drug's known pharmacological profile that is actually
inferable from molecular structure remains systematically underexplored.
A systematic case study using acetylsalicylic acid (ASA, Aspirin) --- one of the
most comprehensively characterized drugs in pharmacology --- serves as model compound.
A Message Passing Neural Network (MPNN) is trained on the Tox21 benchmark and
GNNExplainer is applied to characterize atom-level attribution.
Results indicate that molecular structure explains approximately 45\% (5/11)
of known ASA adverse effects.
A four-category Gap Taxonomy (GAP-1 through GAP-4) is introduced distinguishing
between principally non-encodable effects, data gaps arising from Missing Not At
Random (MNAR) mechanisms, assay panel mismatches, and representation errors.
The MNAR gap is empirically quantified via a systematic ChEMBL query
(42 documented assays, 0 retrievable bioactivity entries).
An attention pooling experiment localizes the representation error
to the MPNN message passing layers rather than the aggregation step.
The Gap Taxonomy has direct implications for drug safety signal detection
and regulatory frameworks including Good Pharmacovigilance Practice
(GVP) guidelines and New Approach Methodologies (NAMs).
Structural limits identified are confirmed in a companion DDI ablation study.
\end{abstract}

\smallskip
\noindent\textbf{Keywords:} Graph Neural Networks, Drug Toxicity, GNNExplainer,
Missing Not At Random, Pharmacovigilance, Tox21, Attention Pooling,
New Approach Methodologies, Gap Taxonomy

\vspace{0.5em}
\noindent\small This work is licensed under
\href{https://creativecommons.org/licenses/by/4.0/}{CC BY 4.0}.

\newpage

\section{Introduction}

Throughout this paper, abbreviations are expanded at first occurrence;
a complete glossary is provided in Table~\ref{tab:glossary}.

\subsection{Motivation}

The prediction of adverse drug reactions (ADR) and drug toxicity from
molecular structure alone represents one of the central challenges in
computational pharmacology.
Early identification of toxic compounds reduces attrition in drug development,
lowers costs, and --- most critically --- protects patients from preventable harm.
Graph Neural Networks (GNNs) have emerged as a structurally natural approach:
molecules are inherently graphs, and GNNs operate directly on atomic connectivity
without the information loss inherent to fixed-length fingerprints or text-based
molecular encodings \citep{Bronstein2017,Duvenaud2015}.

Despite rapid methodological progress, a fundamental question remains
underexplored: what fraction of a drug's known pharmacological profile is
actually encodable in its molecular structure?
This question is not merely academic.
As pharmacovigilance workflows increasingly incorporate machine learning
components, regulatory frameworks are adapting accordingly.
Good Pharmacovigilance Practice (GVP) guidelines --- the primary regulatory
framework governing drug safety monitoring in the European Union (EU) --- are
being extended to address transparency and mechanistic coverage requirements
for AI-based safety systems (European Medicines Agency (EMA) AI Workplan
2023--2028).
New Approach Methodologies (NAMs) --- the regulatory umbrella term for modern
alternatives to animal testing including in-silico, in-vitro, and systems biology
approaches --- are increasingly used for safety assessment.
A key unresolved question for both GVP and NAMs is whether their assay panels
provide sufficient mechanistic coverage for specific drug classes.

Acetylsalicylic acid (ASA, commonly known as Aspirin) is selected as the model
compound for five methodologically motivated reasons.
First, 125 years of clinical use have produced one of the most complete adverse
effect profiles in pharmacology, enabling ground-truth validation at a level
rarely achievable for newer compounds.
Second, both on-target toxicity (cyclooxygenase (COX)-1 inhibition causing
gastrointestinal (GI) bleeding and platelet dysfunction) and off-target toxicity
(mitochondrial uncoupling, hepatotoxicity) are characterized at mechanistic
resolution, providing a stringent test for structure-based prediction.
Third, the adverse effect profile spans all four Gap categories (defined in
Section~\ref{sec:gaps}) --- making ASA uniquely suited to demonstrate the full
taxonomy.
Fourth, mechanisms are understood at atomic resolution, enabling
pharmacophore-level GNNExplainer validation.
Fifth, with only 13 heavy atoms, atomic attribution remains directly
interpretable --- a property lost in larger molecules where GNNExplainer outputs
become difficult to map onto discrete pharmacophores.

\subsection{Related Work}

Graph-based molecular representations for toxicity prediction have been advanced
by Message Passing Neural Networks (MPNNs) \citep{Gilmer2017} and their
application to the Tox21 benchmark (Toxicology in the 21st Century, a public
high-throughput screening dataset with 12 toxicological assay endpoints
\citep{Mayr2016}).
Multi-task learning across Tox21's 12 endpoints exploits shared structural motifs
and has become standard practice \citep{Yang2019}.
Post-hoc explainability via GNNExplainer \citep{Ying2019} reveals which atoms
drive classification decisions, enabling comparison against known pharmacophores.
The Missing Not At Random (MNAR) problem \citep{Rubin1976} --- where data absence
correlates with the missing value itself --- has been recognized in
cheminformatics \citep{Sheridan2013,Wenzel2019} but rarely quantified empirically
for specific endpoints.
Attention pooling \citep{Li2015} offers a potential remedy for atom-weighting
biases in mean pooling, but its effect on pharmacophore attribution has not been
studied against GNNExplainer ground truth.

\subsection{Contributions}
\label{sec:contributions}

This study makes four contributions:

\begin{enumerate}[leftmargin=*]
\item A four-category Gap Taxonomy (GAP-1 through GAP-4) classifying
      failure modes of structure-based toxicity prediction by root cause.
\item Empirical GNNExplainer-based localization of a representation error
      (GAP-4) in ASA: the model focuses on the aromatic ring rather than
      the pharmacologically relevant carboxyl group for mitochondrial
      uncoupling endpoints (SR-MMP pharmacophore ratio: 0.646 under mean
      pooling).
\item Empirical quantification of the MNAR gap (GAP-2) via a systematic
      ChEMBL query \citep{Mendez2019}: 42 documented SR-MMP assays,
      0 retrievable bioactivity entries for carboxylic acid structures ---
      confirming that the data necessary to resolve GAP-4 resides in
      proprietary repositories.
\item An attention pooling experiment demonstrating that GAP-4 originates
      in the MPNN message passing layers, not in the aggregation step:
      the gate network correctly upweights carboxyl atoms ($\alpha=0.62$),
      but atom embeddings remain ring-dominated from the confounded
      training distribution.
\end{enumerate}

\section{Methods}

\subsection{Molecular Graph Construction}

ASA (SMILES: \texttt{CC(=O)Oc1ccccc1C(=O)O} --- a text-based notation
where atoms are letters, bonds are implicit, and rings are denoted by numbers)
is converted to a molecular graph using RDKit \citep{Landrum2006}.
The graph has 13 nodes (heavy atoms) and 26 directed edges (13 bonds, each
represented in both directions for bidirectional message passing).

Each atom is described by a 31-dimensional feature vector:
atom type (11 binary flags), hybridization state (6 flags),
formal charge (1), hydrogen count (5 flags), ring membership (1),
aromaticity (1), chirality (4 flags), and normalized atomic mass (1).
Each bond is described by 12 features: bond type (5 flags), conjugation (1),
ring membership (1), and stereo configuration (5 flags).
Numbers in square brackets throughout refer to the numbered references.

\subsection{The Four Gap Categories}
\label{sec:gaps}

\begin{infobox}{Gap Taxonomy Overview}
\textbf{GAP-1} \textit{Principally non-encodable:} the adverse effect requires
patient-level information (age, genetics, co-infection, dose history) absent from
the molecular graph. No amount of additional training data or better architecture
can resolve this gap.\\[4pt]
\textbf{GAP-2} \textit{Data gap (MNAR):} the effect is structurally predictable
in principle, but training data is systematically absent from public databases ---
not by random chance, but because assays were never performed or results remain
proprietary.\\[4pt]
\textbf{GAP-3} \textit{Assay panel mismatch:} no assay in the training panel
(Tox21) measures the relevant biological mechanism. The model cannot learn what
it was never shown.\\[4pt]
\textbf{GAP-4} \textit{Representation error:} the model learns the wrong
structural feature as the predictor, despite the correct feature being present
in the molecule. Typically caused by confounding in the training data.
\end{infobox}

Two study hypotheses are evaluated:
(H1) GAP-3 (assay panel mismatch) is the dominant failure mode for
Non-Steroidal Anti-Inflammatory Drugs (NSAIDs) on the Tox21 panel;
(H2) GAP-4 for SR-MMP originates in the message passing layers rather than
the aggregation (pooling) step.

\subsection{MPNN Architecture}

The MPNN follows \citep{Gilmer2017} with neural network convolution (NNConv)
as the message passing operator.
For each bond, NNConv computes a bond-specific weight matrix from bond features,
so that single and double bonds produce distinct messages --- chemically
meaningful since bond type determines electron distribution and reactivity.

Architecture: input projection (31d$\to$64d), three NNConv layers (64d, with
BatchNorm, Rectified Linear Unit (ReLU) activation, and Dropout $p=0.2$),
global mean pooling, and a two-layer multi-layer perceptron (MLP) classifier
(424,140 parameters total).
All 12 Tox21 endpoints are trained simultaneously (multi-task learning).
Missing labels are handled via NaN masking: only observed labels contribute to
the Binary Cross-Entropy (BCE) loss.

\subsection{Attention Pooling}

Standard mean pooling weights all atoms equally.
Attention pooling \citep{Li2015} instead learns which atoms matter: a gate
network (MLP: 64d$\to$32d$\to$1) produces an importance score per atom,
normalized via softmax.
The molecule embedding is the weighted sum of atom embeddings.
The gate network adds 2,113 parameters (0.5\% increase).

\subsection{GNNExplainer and Pharmacophore Ratio}
\label{sec:explain}

GNNExplainer \citep{Ying2019} identifies which atoms most influence a specific
prediction by optimizing atom importance masks (200 optimization iterations).
The pharmacophore ratio summarizes the result:
\begin{equation}
    \text{Pharmacophore Ratio} =
    \frac{\text{mean importance of pharmacophore atoms}}
         {\text{mean importance of all other atoms}}
\end{equation}
Thresholds of 1.5 (correct focus) and 1.0 (incorrect focus) are defined
heuristically for exploratory purposes and require calibration against larger
molecular sets in future work.
For SR-MMP (mitochondrial membrane potential), the pharmacophore is the carboxyl
group (atoms 10--12, pKa 3.5), which drives the proton carrier mechanism.

\subsection{Gap Taxonomy Construction and MNAR Quantification}

The Gap Taxonomy is constructed through three-level validation:
(a)~Tox21 endpoint ground truth based on established pharmacology;
(b)~clinical adverse effects documented in DrugBank \citep{Wishart2018} and
SIDER (Side Effect Resource);
(c)~a manually constructed Tox21-to-clinical mapping based on the biochemical
cascade from molecular initiating event to clinical outcome.
DrugBank was accessed in early 2026; a complete download was temporarily
unavailable at the time of writing.

The MNAR gap (GAP-2) is quantified via a ChEMBL \citep{Mendez2019} query.
Structural filters: free carboxyl group (SMARTS: \texttt{[CX3](=O)[OX2H1]}),
logP $> 1.5$ (lipophilicity threshold for membrane permeability),
and molecular weight 100--600 Da.

\section{Results}

\subsection{MPNN Performance on Tox21}

Table~\ref{tab:tox21} summarizes MPNN performance across all 12 Tox21 endpoints.
The mean ROC-AUC (Area Under the ROC Curve) is 0.69.
Stress response (SR-*) endpoints are learned substantially better than nuclear
receptor (NR-*) endpoints.
``Relevant'' indicates a direct mechanistic link to known ASA adverse effects;
``marginal'' indicates a link only under specific conditions; ``none'' means
no established link.

\begin{table}[H]
\centering
\caption{MPNN performance per Tox21 endpoint. AUC = Area Under the ROC Curve
(1.0 = perfect, 0.5 = random). SR: Stress Response. NR: Nuclear Receptor.
MMP: Mitochondrial Membrane Potential. ARE: Antioxidant Response Element.}
\label{tab:tox21}
\small
\begin{tabular}{llcc}
\toprule
\textbf{Endpoint} & \textbf{Full Name} & \textbf{AUC} & \textbf{Relevance for ASA} \\
\midrule
NR-AR       & Androgen Receptor Agonism         & 0.38 & None \\
NR-AR-LBD   & AR Ligand Binding Domain          & 0.59 & None \\
NR-AhR      & Aryl Hydrocarbon Receptor         & 0.80 & Marginal \\
NR-Aromatase& Aromatase (CYP19A1) Inhibition    & 0.22 & None \\
NR-ER       & Estrogen Receptor Agonism         & 0.63 & None \\
NR-ER-LBD   & ER Ligand Binding Domain          & 0.61 & None \\
NR-PPAR-$\gamma$ & PPAR-gamma Agonism           & 0.68 & Marginal \\
SR-ARE      & Antioxidant Response Element      & 0.88 & \textbf{Relevant} \\
SR-ATAD5    & DNA Damage Response Checkpoint    & 0.89 & None \\
SR-HSE      & Heat Shock Element                & 0.82 & Marginal \\
SR-MMP      & Mitochondrial Membrane Potential  & 0.95 & \textbf{Relevant} \\
SR-p53      & p53 Apoptosis Signaling           & 0.82 & Marginal \\
\midrule
\textbf{Mean} & & \textbf{0.69} & \\
\bottomrule
\end{tabular}
\end{table}
\vspace{0.8em}

\subsection{ASA Inference}

ASA is not included in the Tox21 training set, making all inference results
genuine zero-shot predictions on an unseen molecule. The pharmacophore
attributions therefore reflect generalization behavior, not memorization of
ASA-specific training signal.

When the trained model evaluates ASA, all 12 Tox21 endpoints are classified as
negative (prediction below the 0.5 decision threshold, derived from the sigmoid
of the raw model output).
SR-ARE ($P = 0.130$) and SR-MMP ($P = 0.069$) produce the highest probabilities.
Both are pharmacologically relevant false negatives: the model predicts no
toxicity signal for endpoints where salicylate effects are well-established.
This is the central finding --- not a contradiction, but evidence that the
failure resides in assay design (GAP-3) and data confounding (GAP-4), not
absence of biological effect.

\subsection{GNNExplainer Atom Attribution}
\label{sec:explainer}

Table~\ref{tab:ratio} shows pharmacophore ratios for selected endpoints.
For SR-MMP, the ratio under mean pooling is 0.646 --- below 1.0, indicating
focus on the aromatic ring (atoms 4--9) rather than the carboxyl group
(atoms 10--12).
The carboxyl group drives the proton carrier mechanism (pKa 3.5): at
mitochondrial matrix pH, the protonated acid diffuses across the inner
membrane, releases the proton, and returns as anion --- dissipating the
electrochemical gradient.
This mechanism is invisible to a model that has learned to associate
aromaticity with SR-MMP positivity from polycyclic aromatic hydrocarbons (PAH)
and quinones in the Tox21 training set.
Attention pooling improves the SR-MMP ratio from 0.646 to 0.805, but does not
reach the correct-focus threshold.
NR-AhR serves as a positive control (ratio: 1.38): the aromatic ring is
correctly identified as the pharmacophore for aryl hydrocarbon receptor
activation.

\begin{table}[H]
\centering
\caption{GNNExplainer pharmacophore ratio for selected Tox21 endpoints applied
to ASA. Ratio $> 1.5$: correct focus. Ratio $< 1.0$: incorrect focus.
MMP: Mitochondrial Membrane Potential. ARE: Antioxidant Response Element.
AhR: Aryl Hydrocarbon Receptor. NR: Nuclear Receptor.}
\label{tab:ratio}
\small
\begin{tabular}{llccc}
\toprule
\textbf{Endpoint} & \textbf{Expected pharmacophore} &
\textbf{Mean Pool} & \textbf{Att Pool} & \textbf{Verdict} \\
\midrule
SR-MMP & Carboxyl group (pKa 3.5) & 0.646 & 0.805 & GAP-4 partial \\
SR-ARE & Carboxyl group (Nrf2)    & 0.835 & 0.766 & GAP-4, no improvement \\
NR-AhR & Aromatic ring (AhR)     & 1.38  & n/a   & Correct (control) \\
NR-AR  & None (no pharmacophore) & n/a   & n/a   & GAP-3 \\
\bottomrule
\end{tabular}
\end{table}
\vspace{0.8em}

\subsection{Gap Taxonomy}

\begin{figure}[H]
\centering
\includegraphics[width=\textwidth]{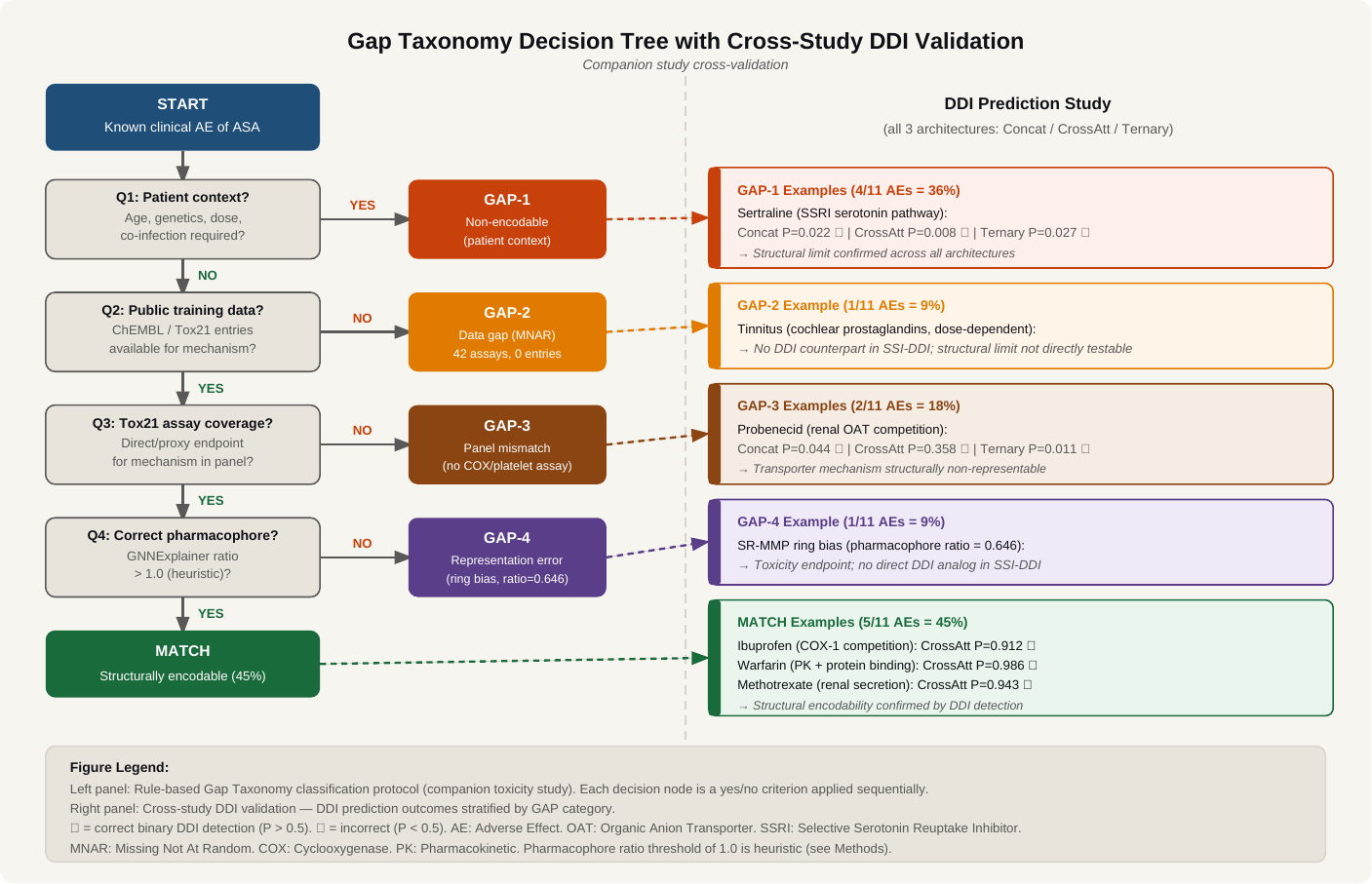}
\caption{Gap Taxonomy decision tree (left) with cross-paper DDI
validation (right, from the companion DDI study \citep{DieterichB2026}).
GAP-1 and GAP-3 structural limits are confirmed by consistent prediction
failure across all three architectures.
MATCH cases are correctly detected.
AE: Adverse Effect. OAT: Organic Anion Transporter.
SSRI: Selective Serotonin Reuptake Inhibitor.}
\label{fig:taxonomy}
\end{figure}

Figure~\ref{fig:taxonomy} presents the operational classification protocol as a decision tree (left panel) with cross-paper validation from the companion DDI study \citep{DieterichB2026} (right panel).
The four sequential criteria (Q1--Q4) are sufficient to reproduce the GAP classification for any drug with a known adverse effect profile.
The right panel provides independent empirical validation: GAP-1 (serotonin pathway, sertraline) and GAP-3 (renal OAT competition, probenecid) structural limits predicted by the taxonomy are confirmed by consistent DDI prediction failure across all three GNN architectures tested in the companion study --- Concat, CrossAtt, and Ternary.
MATCH cases (ibuprofen, warfarin) are correctly predicted, consistent with their structural encodability.
This cross-paper triangulation supports the central claim: structural limits in toxicity prediction transfer directly to DDI prediction, independent of architectural complexity.

Table~\ref{tab:gaps} presents the full Gap Taxonomy for ASA.
An important distinction applies to the MATCH category: it denotes effects that
are structurally encodable in principle (the molecular mechanism is present in
the molecular graph), not necessarily that this model predicted them correctly.
For example, GI hemorrhage is MATCH because COX-1 inhibition is encoded in the
carboxyl/ester structure --- but no Tox21 assay measures it (GAP-3 at the assay
level), and the model does not predict it positively.
GAP-3 (assay panel mismatch) dominates Tox21 endpoint failures (6/12 = 50\%),
supporting H1.
For clinical adverse effects, 45\% (5/11) are structurally encodable (MATCH).

\begin{table}[H]
\centering
\caption{Gap Taxonomy for ASA. Each gap type classifies a distinct root cause
for prediction failure. MNAR: Missing Not At Random. GI: Gastrointestinal.
COX: Cyclooxygenase. AEs: Adverse Effects.}
\label{tab:gaps}
\small
\begin{tabular}{p{1.5cm}p{3.5cm}p{2cm}p{3cm}p{3.5cm}}
\toprule
\textbf{GAP} & \textbf{Definition} & \textbf{Tox21} &
\textbf{Clinical AEs} & \textbf{Resolution} \\
\midrule
GAP-1 & Non-encodable: requires patient context &
0 (0\%) & 4 (36\%): Reye, poisoning, fetal, allergic &
Not resolvable with structure models \\
GAP-2 & Data gap (MNAR): data absent from public databases &
0 (0\%) & 1 (9\%): Tinnitus &
Federated learning; targeted assays \\
GAP-3 & Panel mismatch: Tox21 does not measure the mechanism &
6 (50\%) & 2 (18\%): GI bleeding, platelet inhibition &
COX-specific assay panel \\
GAP-4 & Representation error: wrong atomic attribution &
2 (17\%) & 1 (9\%): Hepatotoxicity &
Data augmentation with lipophilic acids \\
MATCH & Structurally encodable in principle &
4 (33\%) & 5 (45\%): GI hemorrhage, nausea, dyspepsia, platelets, bronchospasm &
No intervention required \\
\bottomrule
\end{tabular}
\end{table}
\vspace{0.8em}

\subsection{MNAR Quantification}

A ChEMBL \citep{Mendez2019} query for SR-MMP-active carboxylic acids yielded
42 documented assays, but 0 retrievable bioactivity entries after applying the
structural filter.
Only 3 curated literature molecules matched all criteria --- far below the
minimum of approximately 100 molecules needed for model retraining.
This finding is consistent with an MNAR-like public availability bias: the
necessary training data appears underrepresented in public databases, consistent
with its location in proprietary Drug Metabolism and Pharmacokinetics (DMPK)
repositories.
Alternative explanations including query-definition mismatch cannot be excluded;
the finding should be interpreted as indicative rather than conclusive evidence
of MNAR.

\subsection{Attention Pooling Experiment}

The attention gate network correctly upweights carboxyl group atoms
(atoms 11 and 12, combined $\alpha = 0.62$), while aromatic ring atoms receive
lower weights ($\alpha = 0.01$--$0.06$ per atom).
Despite this correct pooling behavior, the SR-MMP pharmacophore ratio improves
only from 0.646 to 0.805.
This localizes GAP-4 to the NNConv message passing layers: atom embeddings
produced by three rounds of message passing are already ring-dominated before
pooling occurs.
H2 is supported by the available evidence: the attention pooling experiment is consistent with a message-passing origin of the representation error, though definitive confirmation would require additional experiments across multiple molecular series.

\section{Discussion}

\subsection{Implications for Drug Safety Signal Detection}

A 50\% panel mismatch for NSAIDs is not a minor technical limitation --- it is
a structural blind spot with direct consequences for automated safety systems
and pharmacovigilance workflows.

\subsubsection*{Mechanistic undercoverage in automated screening}

Many automated High-Throughput Screening (HTS) and AI-based toxicity systems
implicitly assume that assay coverage approximates biological risk coverage.
For NSAIDs, this assumption fails.
NSAID risks arise primarily from on-target pharmacology: COX-1 inhibition is
simultaneously the therapeutic mechanism and the cause of the most clinically
significant adverse effects.
Tox21 was designed for generic stress responses, nuclear receptor interactions,
and genotoxicity --- not for functional pharmacology, hemodynamic effects, or
tissue-level physiology.

Platelet inhibition via COX-1/prostaglandin-endoperoxide synthase 1 (PTGS1) and
thromboxane A2 (TXA2) suppression involves kernelless, metabolically specialized
cells absent from standard cell line assays.
GI bleeding arises from the combined loss of protective prostaglandins in the GI
mucosa and simultaneous platelet dysfunction --- a coupled, tissue-level process
requiring multilayer barrier models, microcirculation, and chronic exposure
conditions that Tox21 assays cannot represent.

\subsubsection*{Signal detection and pharmacovigilance}

In pharmacovigilance practice, safety signal detection relies on triangulation
across multiple evidence streams: disproportionality analysis, mechanistic
biological plausibility, preclinical data, and class effects.
When preclinical mechanistic data is absent because the assay panel does not
cover the relevant mechanism, one triangulation pillar is structurally weakened.

This has regulatory implications under GVP Module IX (Signal Management), which
emphasizes multiple evidence streams and biological plausibility.
GVP Module V (Risk Management Systems) requires identification of missing
information and mechanistic uncertainty.
A 50\% panel mismatch constitutes a quantifiable limitation of nonclinical
safety coverage.
GVP Module VIII (Post-Authorisation Safety Studies) becomes more critical when
preclinical models are insufficient.

\subsubsection*{New Approach Methodologies and NAM validation}

The findings are directly relevant to ongoing regulatory discussions around NAMs.
A central question in NAM qualification is whether a platform provides
fit-for-purpose mechanistic coverage.
The Gap Taxonomy addresses this: assay count alone does not equal biological
coverage.
Tox21's 12 endpoints constitute a large panel by historical standards, yet 50\%
of NSAID-relevant mechanisms are unrepresented.
Future NAM qualification frameworks should require mechanism-specific coverage
assessments for each drug class of interest.

Concretely, resolution of GAP-3 for NSAIDs would require: COX-1/PTGS1 functional
assays, TXA2 measurement, platelet aggregation systems, GI mucosal barrier
models (organoids or transwell), and Adverse Outcome Pathway (AOP) --- structured
representations of toxicological mechanisms from molecular initiating event to
adverse outcome --- based integration.
Until such extensions are available, mechanistic absence in HTS/NAM data must
not be interpreted as a negative safety signal for NSAID-class compounds.

\subsection{The MNAR Problem in Public Pharmacological Databases}

The ChEMBL finding --- 42 assays, 0 retrievable entries --- illustrates that the
MNAR problem extends beyond label imbalance.
Public databases systematically underrepresent proprietary DMPK screening data,
particularly for niche mechanistic endpoints such as mitochondrial uncoupling by
lipophilic carboxylic acids.
Regulatory pharmacovigilance databases offer complementary signal sources:
the FDA Adverse Event Reporting System (FAERS, U.S. Food and Drug Administration (FDA)),
EudraVigilance (EMA), and WHO VigiBase (30+ million individual case safety
reports from 130+ countries).
Federated learning frameworks \citep{Oldenhof2023} could provide industry data
access without proprietary disclosure.

\subsection{GAP-4: Representation Error vs.\ Data Problem}

The attention pooling experiment provides mechanistic insight: GAP-4 originates
in the NNConv message passing layers.
The gate network correctly identifies carboxyl atoms as important
(combined $\alpha = 0.62$), but three rounds of message passing have already
encoded ring-dominated atom embeddings from the confounded Tox21 distribution.
The fix requires either mechanistically-informed message passing (explicitly
encoding functional group properties) or training data that breaks the
PAH/carboxylic acid confound --- which requires resolving GAP-2 first.

\subsection{Limitations}

The study uses a subset of 500 molecules from the Tox21 benchmark
($\approx$7,831 compounds total \citep{Mayr2016}), limiting statistical power
for endpoint-specific analyses.
ASA is a single model compound --- generalizability requires validation across
other drug classes.
The 3D conformation is not encoded: stereoisomers and conformer-dependent
interactions are invisible to the 2D graph.
The curated reference set of 7 drug pairs is too small for statistically robust
threshold optimization.
DrugBank 5.1.20 was temporarily unavailable for complete download at the time
of writing; the Tox21-to-clinical mapping relied on curated literature entries.
The ChEMBL MNAR quantification reflects public database status as of May 2026.

\subsection{Future Directions}

The structural limits identified by this taxonomy --- particularly GAP-1
(principally non-inferable effects requiring patient context) and GAP-3
(assay panel mismatch for COX-dependent pharmacology) --- are amenable to
cross-validation in drug-drug interaction prediction settings.
In such settings, the same mechanistic constraints should produce consistent
prediction failures independent of architectural complexity:
serotonin pathway-mediated interactions (GAP-1) and renal transporter
competition (GAP-3) should fail across all structure-based architectures.
Such cross-paper validation is reported in a companion DDI ablation study
\citep{DieterichB2026}, where GAP-1 and GAP-3 structural limits are confirmed
by consistent prediction failure across all three GNN architectures tested.

\section{Conclusion}

This study demonstrates that molecular structure explains approximately 45\%
(5/11) of the known adverse effect profile of acetylsalicylic acid when using
a state-of-the-art MPNN trained on Tox21.
The remaining 55\% distributes as: GAP-1 (36\%, 4/11), GAP-2 (9\%, 1/11),
GAP-3 (0\% of clinical AEs --- affects Tox21 panel coverage only),
and GAP-4 (9\%, 1/11).

The Gap Taxonomy provides an operational framework for characterizing
structural information limits that goes beyond aggregate AUC metrics.
A missing mechanism cannot be inferred by better algorithms --- it is a
structural information problem, not a model deficiency.
This distinction has direct consequences for how prediction failures
should be interpreted in structure-based toxicity and interaction prediction systems.
The MNAR quantification --- 42 SR-MMP assays, 0 retrievable entries ---
demonstrates that GAP-4 cannot be resolved through better architecture alone.
The attention pooling experiment confirms that the representation error resides
in the message passing layers, not in aggregation.

Three directions emerge:
(1)~COX-specific and platelet-function assay panels for NSAID coverage (GAP-3);
(2)~federated learning with pharmaceutical manufacturers for DMPK data access
(GAP-2);
(3)~integration of pharmacovigilance databases (FAERS, EudraVigilance, WHO
VigiBase) for postmarketing signal enrichment (GAP-1 and GAP-2).
The methodology --- gap classification plus empirical MNAR quantification via
database queries --- is transferable to other GNN-based toxicity and interaction
prediction systems and other compound classes.

\section*{Conflicts of Interest}

The author declares no conflicts of interest. No external funding was received
for this study. The author has no financial relationships with any organization
that might have an interest in the submitted work.

\section*{Data and Code Availability}

The Gap Taxonomy annotation for ASA (\texttt{aspirin\_ground\_truth.json}) is
available from the author upon reasonable request.
Model scripts and trained weights are available upon reasonable request.
The Tox21 training data is publicly available via the chemprop
repository~\citep{Yang2019}.
ChEMBL data were accessed via the \texttt{chembl-webresource-client} v0.10.9
in May 2026.
This work is licensed under CC BY 4.0
(\url{creativecommons.org/licenses/by/4.0}).

\clearpage
\section*{Glossary of Abbreviations}

\begin{table}[H]
\centering
\caption{Abbreviations used in this paper, listed alphabetically. INN: International Nonproprietary Name.}
\label{tab:glossary}
\small
\begin{tabular}{ll}
\toprule
\textbf{Abbreviation} & \textbf{Full Term} \\
\midrule
ADR    & Adverse Drug Reaction \\
AhR    & Aryl Hydrocarbon Receptor \\
AOP    & Adverse Outcome Pathway \\
ASA    & Acetylsalicylic Acid (Aspirin; INN: ASA) \\
AUC    & Area Under the ROC Curve \\
BCE    & Binary Cross-Entropy (loss function) \\
ChEMBL & Chemical biological activity database (EMBL-EBI) \\
COX    & Cyclooxygenase (COX-1 = PTGS1, COX-2 = PTGS2) \\
DMPK   & Drug Metabolism and Pharmacokinetics \\
EMA    & European Medicines Agency \\
FAERS  & FDA Adverse Event Reporting System \\
FDA    & U.S. Food and Drug Administration \\
GI     & Gastrointestinal \\
GNN    & Graph Neural Network \\
GVP    & Good Pharmacovigilance Practice (EMA guidelines) \\
HTS    & High-Throughput Screening \\
iPSC   & Induced Pluripotent Stem Cell \\
logP   & log Octanol-Water Partition Coefficient (lipophilicity) \\
MLP    & Multi-Layer Perceptron \\
MNAR   & Missing Not At Random \\
MPNN   & Message Passing Neural Network \\
NAM    & New Approach Methodology \\
NNConv & Neural Network Convolution \\
NR     & Nuclear Receptor (Tox21 endpoint prefix) \\
NSAID  & Non-Steroidal Anti-Inflammatory Drug \\
PAH    & Polycyclic Aromatic Hydrocarbon \\
PBPK   & Physiologically Based Pharmacokinetics \\
pKa    & Negative log of acid dissociation constant \\
PTGS1  & Prostaglandin-Endoperoxide Synthase 1 (= COX-1) \\
QSAR   & Quantitative Structure-Activity Relationship \\
RDKit  & Open-Source Cheminformatics Library \\
ReLU   & Rectified Linear Unit (activation function: max(0,x)) \\
ROC    & Receiver Operating Characteristic \\
ROS    & Reactive Oxygen Species \\
SIDER  & Side Effect Resource \\
SMILES & Simplified Molecular Input Line Entry System \\
SR     & Stress Response (Tox21 endpoint prefix) \\
Tox21  & Toxicology in the 21st Century (NIH/EPA/NCATS initiative) \\
TXA2   & Thromboxane A2 \\
\bottomrule
\end{tabular}
\end{table}

\clearpage
\bibliographystyle{unsrt}
\bibliography{what_molecular_structure_cannot_tell_us}

\end{document}